\documentclass[]{ceurart}

\usepackage{subfigure}
\usepackage{amsmath}
\usepackage{natbib}
\usepackage{amsfonts}
\usepackage{graphicx}
\usepackage{hhline}
\usepackage{hyperref}
\usepackage{array}
\usepackage{multicol}
\usepackage{multirow}
\usepackage{adjustbox}

\usepackage{marginnote}
\newcommand{\pageenlarge}[1]{}

\newcommand{\wtu}{\triangle}

\newcommand{\btu}{\blacktriangle}

\newcommand\boldparagraph[1]{\vspace{0.6em}\noindent\textbf{#1}}

\newcommand{\iadh}[1]{\textcolor{black}{#1}}
\newcommand{\craig}[1]{\textcolor{black}{#1}}
\newcommand{\nic}[1]{\textcolor{black}{#1}}

\newcommand{\craigi}[1]{\textcolor{black}{#1}}
\newcommand{\crc}[1]{\textcolor{black}{#1}}

\begin{document}

\copyrightyear{2021}
\copyrightclause{Copyright for this paper by its authors.
  Use permitted under Creative Commons License Attribution 4.0
  International (CC BY 4.0).}

\conference{IIR 2021: The 11th Italian Information Retrieval Workshop, September 13--15, 2021, Bari, Italy}

\title{On Single and Multiple Representations in Dense Passage Retrieval}
\title[mode=sub]{(Full original paper)}

\author[1]{Craig Macdonald}
\author[2]{Nicola Tonellotto}
\author[1]{Iadh Ounis}

\address[1]{University of Glasgow, UK}
\address[2]{University of Pisa, Italy}

\maketitle

\begin{abstract}
The advent of contextualised language models has brought gains in search effectiveness, not just when applied for re-ranking the output of classical weighting models such as BM25, but also \iadh{when} used directly for passage indexing and retrieval, a technique which is called dense retrieval. \iadh{In the} existing literature in neural ranking, two dense retrieval families have become apparent: single \craig{representation}, where entire passages are represented by a single embedding (usually BERT's [CLS] token, as exemplified by the recent ANCE approach), or multiple \iadh{representations}, where each token in a passage is represented by its own embedding (as exemplified by the recent ColBERT approach). 
\iadh{These two families} have not been directly compared. \iadh{However, because of} the likely importance of dense retrieval moving forward, a clear understanding of their advantages and disadvantages is \craig{paramount}. To this end, this paper contributes a direct study on their comparative effectiveness, noting situations where each \iadh{method} under/over performs w.r.t. each other, and w.r.t.\ a BM25 baseline. 
\nic{We observe that, while ANCE is more efficient than ColBERT in terms of response time and memory usage, multiple representations are statistically more effective than the single representations for MAP and MRR@10. We \iadh{also} show that multiple representations get better improvements than single representations for queries being \iadh{the} hardest for BM25, \iadh{as well as for} definitional queries, and those with complex information needs.}
\end{abstract}

\section{Introduction}
\pageenlarge{2}
\looseness -1 Pre-trained contextualised language models such as BERT have been shown to greatly improve \iadh{retrieval} effectiveness over the \craig{previous} \iadh{state-of-the-art} methods in many information retrieval \iadh{(IR)} tasks~\cite{bert}. These contextualised language models are able to learn semantic representations called \emph{embeddings} from the contexts of words \iadh{and}, \craig{therefore,} better capture the relevance of a document w.r.t.\ a query, with substantial improvements over the classical approach in the ranking and re-ranking of documents~\cite{lin2020pretrained}. 
Most BERT-based models are computationally expensive \craig{\iadh{for estimating} query-document similarities \iadh{in} ranking, due to the complexity of the underlying transformer neural network}~\cite{khattab2020colbert,Hofsttter2019LetsMR,snrm}. As such, BERT-\craig{based} ranking models have been used as second-stage rankers in retrieval cascades, in particular to re-rank candidate documents generated by classical relevance models such as BM25~\cite{cedr,prettr,epic}. \craig{BERT-based models are also limited in the length of text that they can process, and hence are often applied on passages rather than full documents (which we focus on in this paper); entire document rankings can be obtained by estimating relevance at a passage level, then aggregating~\cite{dai2019deeper}.}

\pageenlarge{2} Recently, several works \craig{have proposed} investigating \iadh{whether} BERT-based systems \iadh{are} able to identify the relevant passages \craig{among} all passages in a collection, \craig{rather than just among} a query-dependent sample; these systems represent a new \craig{type of retrieval approaches} called \emph{dense retrieval}. \nic{In dense retrieval, passages are represented by real-valued vectors, \iadh{while the} query-document similarity is computed by deploying efficient nearest neighbour techniques over specialised indexes, such as \craig{those provided by the FAISS toolkit}~\cite{JDH17}}. 
Thus far, two different families of dense retrieval approaches have emerged recently, based on {\em single representation} and {\em multiple representation}. In particular, DPR~\cite{karpukhin2020dense} and ANCE~\cite{xiong2020approximate} use a single representation, by indexing only the embedding of BERT's [CLS] token, and therefore \craig{\iadh{this is assumed to} represent the meaning of} an entire passage within that single embedding. At retrieval time, the [CLS] embedding of the query is \craig{then} used to retrieve passages by identifying nearest neighbours using a FAISS index. In contrast, ColBERT~\cite{khattab2020colbert}, \iadh{which uses multiple representation}, indexes an embedding for each token in each document. At retrieval time, a set of the nearest document embeddings to each query embedding is retrieved, by identifying the approximate nearest neighbours from a FAISS index. These passages must then be exactly scored, based on \iadh{the} maximal similarity between \iadh{the} query and \iadh{the passage} embeddings, to obtain the final ranking.

These are two markedly different families of \craig{dense retrieval} approaches. Indeed, as ColBERT records one embedding for every token, this makes for a large index of embeddings, which may allow a richer semantic representation of the content. 
On the other hand, DPR and ANCE rely on a single embedding \iadh{sufficiently} representing the content of each passage. However, at the time of writing, no systematic study has compared these two families of dense retrieval approaches.  For this reason, this work contributes \nic{a first investigation \iadh{into} the effectiveness of single and multiple representation embeddings for dense retrieval, as exemplified by ANCE and ColBERT, respectively. We perform experiments in a controlled environment using the same collection and \iadh{query sets}, and we report several effectiveness metrics, together with a detailed comparison of the results obtained for the two representation families w.r.t. a common baseline, namely BM25. \iadh{To derive further insights, we also} provide a per-query analysis of the effectiveness of single and multiple representations.} \nic{We observe that, while ANCE is more efficient than ColBERT in terms of response time and memory usage, multiple representations are statistically more effective than the single representations for MAP and MRR@10. We \iadh{also} show that multiple representations obtain better improvements than single representations for queries that are \iadh{the} hardest for BM25, \iadh{as well as for} definitional queries, and those with complex information needs.}

\section{Problem Statement} 

\boldparagraph{Embeddings.}
\nic{Contextualized language models such as BERT \craig{have been} trained on a large corpus \craig{for language understanding}, and then fine-tuned on smaller, more specific textual collections targeting a particular IR task. Through this fine-tuning, BERT learns how to map texts, either queries or documents, into a multi-dimensional space of one or more vectors, called \emph{embeddings}. Both queries and documents are tokenised into terms according to a predefined vocabulary; BERT learns a function mapping tokens in a query into multiple query embeddings, one per query term,  and another potentially different function mapping tokens in a document into a document embedding per term. BERT and \iadh{its} derived models \craig{also} \iadh{make use of} special tokens, such as the [CLS] (classification) token, the [SEP] (separator) token, and the [MASK] (masked) token. In particular, [CLS] is always placed at the beginning of any text given as input to BERT, both at training and inference time, and \iadh{is} used to let BERT learn a global representation of the input text as a single embedding. In more detail, a text composed by $m$ terms given as input to BERT will produce $m+1$ embeddings, one per input term plus one additional embedding for [CLS]. In \emph{single representation} models, such as ANCE, the embedding corresponding to [CLS] is assumed to encode all possible information about the input text, including the possible semantic context of the composing terms. In contrast, in \emph{multiple representation} models, such as ColBERT, \craig{each input term's embedding encodes its specific semantic information within the context of the entire input text.}}

\pageenlarge{2} \boldparagraph{Dense Retrieval.}
\looseness -1 \nic{The embeddings produced \iadh{by the} BERT models have recently demonstrated \iadh{their} promise in being a suitable basis for {\em dense retrieval}. In dense retrieval, documents and queries are \iadh{represented} using embeddings. The embeddings from the documents in a collection can be pre-computed through the application of the BERT learned mapping and stored into an index data structure for embeddings supporting \craig{nearest neighbour} similarity search, as exemplified by the FAISS toolkit~\cite{JDH17}.
Depending on the number and dimensions of the embeddings \iadh{stored into the index}, advanced compression strategies, together with suitable nearest neighbour search algorithms, can be employed. In order to reduce the time required to identify the most similar document embeddings to a given input embedding, it is possible to shift from exact nearest neighbour search to approximate nearest neighbour search. While ANCE stores embeddings in an uncompressed format supporting exact search, ColBERT, given the larger number of document embeddings \iadh{it has to} store, \iadh{resorts} to compressed and quantised embeddings supporting approximate search. However, the approximate similarity scores produced by approximate search \crc{are not used by the ColBERT implementation} to compute \iadh{the} final top documents to \iadh{return} for a given query~\cite{approxScoresCIKM}\footnote{\crc{Indeed, in~\cite{approxScoresCIKM} we show that these approximate scores can allow a high recall but low precision ranking to be obtained, which can be used to apply rank cutoffs to the the candidate set.}}.
Hence, ColBERT uses approximate search over compressed \iadh{embeddings} to identify a candidate {\em set} of documents, which are then re-scored using an index with direct lookup for retrieving the candidate documents' embeddings, to obtain the final ranking of documents returned to the user.} 


\boldparagraph{Research \iadh{Questions}.}
\nic{In this work, we aim \iadh{to compare} the single and the multiple embedding representations leveraging the ANCE and ColBERT implementations. \craig{Indeed, there was not an effectiveness comparison with ColBERT in the ANCE paper~\cite{xiong2020approximate}. \iadh{ANCE} embodies a recent} single representation approach, \iadh{where we have} a single large embedding per query/document, \iadh{which} can be processed with exact similarity search in a single stage. In contrast, in the multiple representation approach (ColBERT), we have a smaller sized embedding for each term in \iadh{the} queries/documents, but due to the large number of embeddings, they must be processed using approximate similarity search. Thereafter the candidate set  must be re-ranked to \iadh{compute} the exact similarity scores.} \craig{The need \iadh{for} ColBERT to re-score all documents in the candidate set necessitates storing all document embeddings in memory. As noted by Lin et al.~\cite{lin2020pretrained}, this presents a significant storage overhead. This underlines the importance of \iadh{an} in-depth analysis of the pros and cons of both approaches, in particular:}
\begin{itemize}
    \item \textbf{RQ1.} What is the effectiveness of single and multiple representations in dense retrieval, in terms of MAP, NDCG@10, MRR@10? 
    \item \textbf{RQ2.} What are the relative gains and losses of single and multiple representations w.r.t. a common baseline such as BM25?
    \item \textbf{RQ3.} For which queries are single representations better than multiple representations, and vice-versa?
\end{itemize}
\looseness -1 In Section~\ref{sec:exps}, we perform comparative experiments to address these research questions.

\section{Experiments}\label{sec:exps}
In the following, we report our experimental setup, followed by analyses for RQs 1-3.
\subsection{Setup}

Our experiments use the MSMARCO passage ranking dataset, \crc{a dataset of 8.8M passages} \crc{and build upon our PyTerrier IR experimentation platform~\cite{macdonald2020declarative,pyterrierCIKM}}. We \crc{adapt} the ANCE implementation\footnote{\url{https://github.com/microsoft/ANCE}} and the ColBERT implementation\footnote{\url{https://github.com/stanford-futuredata/ColBERT/tree/v0.2}} provided by \iadh{their respective} authors, \crc{using integrations with PyTerrier}\footnote{See \url{https://github.com/terrierteam/pyterrier_ance} and  \url{https://github.com/terrierteam/pyterrier_colbert}.}.  We use the \nic{provided} ANCE model for \iadh{the} MSMARCO passage ranking dataset. We train ColBERT using the same MSMARCO passage ranking training triples file for 44,500 batches. In particular, we follow~\cite{xiong2020approximate}~and~\cite{khattab2020colbert} for the \crc{settings} of ANCE and ColBERT, as summarised in Table~\ref{tab:summary}. 

\looseness -1 \craigi{Of note, while ColBERT fine-tunes the {\tt bert-base-uncased} BERT model, ANCE fine-tunes a RoBERTa model~\cite{liu2019roberta} (specifically {\tt roberta-base}), which is reported to apply more refined pre-training than BERT. To try to eliminate model choice as a confounding factory, we also trained a version of ColBERT by fine-tuning {\tt roberta-base}. We found that even after training for 300k batches (6$\times$ longer than we trained ColBERT using  {\tt bert-base-uncased}), this latter model could had relative performance 25\% less than the BERT-based ColBERT model (around NDCG@10 of 0.533). Hence we discarded the RoBERTa-based ColBERT model from further consideration. On the other hand, all of the released ANCE models use RoBERTa; training ANCE requires multiple GPUs, e.g., 16, and has not, to the best of our knowledge, been reproduced. Hence, we argue that as the RoBERTa-based ANCE and BERT-based ColBERT are individually shown to be effective by their respective authors, the comparison of these representative models still allows for interesting observations.}

%
%
%

\pageenlarge{2} \looseness -1 We index the corpus using the code provided by the authors. 
\craig{Table~\ref{tab:summary} reports the statistics of the resulting indices. In particular, }
the ANCE document index is stored in FAISS using the uncompressed \textsf{IndexFlatIP} format. The ColBERT document index is stored in FAISS using the compressed and quantised \textsf{IndexIVFPQ} format, which is trained on a random 5\% sample of the document embeddings. \craigi{Mean response times for both ANCE and ColBERT, and their memory consumption, are also shown in Table~\ref{tab:summary}.}


\looseness -1 For evaluating effectiveness, we use the publicly available \iadh{query sets} with relevance assessments: 5000 queries sampled from the MSMARCO Dev set -- which contain on average 1.1 judgements per query -- as well as the TREC 2019 \iadh{query set}, which contains 43 queries with an average of 215.3 judgements per query. To measure effectiveness, we employ MRR@10 for the MSMARCO Dev set\footnote{This is the metric recommended by the track organisers for this query set.}, and the MRR@10, NDCG@10 and MAP for the TREC \iadh{query set}. 


To examine gains and losses, for each query and each effectiveness metric,  we examine the comparative reward (improvement) and risk (degradation) over a BM25 baseline (following~\cite{10.1145/2348283.2348385}), as well as \iadh{the} number of wins \& losses (\iadh{improved} and degraded queries). 

\begin{table}[tb]
    \centering
    \caption{Salient statistics of the ANCE and ColBERT \iadh{setups}.}\label{tab:summary}
    \begin{adjustbox}{max width=\textwidth}
    \begin{tabular}{lcc}
        \toprule
         & ANCE & ColBERT \\
        \midrule
        Representation & single & multiple \\
        Base model & {\tt roberta-base} & {\tt bert-base-uncased } \\
        \# emb. per query & 1 & 32 \\
        \# emb. per passage & 1 & up to 180 \\
        Emb. dimensions & 768  & 128 \\
        FAISS index size & 26GB & 16GB \\
        Embedding index size & -- & 176GB\\
        \midrule
        Mean Response Time & 211ms & 635ms\\
        \bottomrule
    \end{tabular}
    \end{adjustbox}
\end{table}

\subsection{Overall Comparison}
\looseness -1 Table~\ref{tab:res1} reports the effectiveness metrics of BM25, ANCE and ColBERT computed on the TREC 2019 and the sample of the MSMARCO Dev \iadh{query sets}. As \craig{expected}, both \iadh{the} ANCE and ColBERT dense retrieval approaches are significantly better than BM25 for \iadh{the} \craig{NDCG@10 \iadh{and} MRR@10 metrics on both query sets.}
Comparing \iadh{the} two dense retrieval approaches, for MAP, ColBERT significantly outperforms ANCE; for NDCG@10, ColBERT enhances ANCE by 6\% (0.6537$\rightarrow$0.6934), but not significantly so; for MRR@10, ANCE is slightly (but not \iadh{significantly}) better than ColBERT on the TREC2019 \iadh{query set} while ColBERT is statistically better than ANCE on MSMARCO Dev by +7\%. \craig{Overall, for RQ1, we conclude} that multiple representations, employed by ColBERT, experimentally obtain better effectiveness than single representations (as employed by ANCE), exhibiting significant boost in effectiveness for MAP (TREC 2019) and MRR@10 (Dev). Among the most striking differences is \iadh{that} for MAP on TREC 2019, where ColBERT markedly outperforms ANCE (and BM25); this observation suggests that the single representation is not sufficiently good at attaining high recall.



\begin{table}[tb]
\caption{Effectiveness metrics of BM25, ANCE and ColBERT on different \iadh{query sets}. Points marked with $\wtu$ and $\btu$ denote a significant increase in effectiveness compared to BM25 and ANCE, respectively, according to a paired t-test with Bonferroni correction (p-value $< 0.05$). }\label{tab:res1}
\begin{adjustbox}{max width=\columnwidth}
\begin{tabular}{lcccc}
\toprule
        & \multicolumn{3}{c}{TREC 2019} & MSMARCO Dev \\
\cmidrule(lr){2-4}\cmidrule(lr){5-5}
        & MAP              & NDCG@10 & MRR@10                 & MRR@10 \\
\midrule
BM25    & 0.2864              & 0.4795 & 0.6410 & 0.1836 \\
ANCE    & 0.3715$^\wtu$       & 0.6537$^\wtu$ & 0.8574$^\wtu$ & 0.3292$^\wtu$ \\
ColBERT & 0.4309$^{\btu\wtu}$ & 0.6934$^\wtu$ & 0.8527$^\wtu$ & 0.3519$^{\wtu\btu}$ \\
\bottomrule
\end{tabular}
\end{adjustbox}\vspace{-\baselineskip}
\end{table}


\subsection{Comparison using a \iadh{Common Baseline}}

\pageenlarge{2}\looseness -1 Next, we investigate the comparative effectiveness of ANCE and ColBERT from the perspective of using BM25 as the reference point, \craig{going further than reporting average performances over \iadh{the} entire \iadh{query sets as} reported in Table~\ref{tab:res1}}. 
To perform this analysis, we define the difficulty of a query according to an effectiveness metric on the BM25 baseline, following~\citet{8919433}.
%
%
Due to the sparsity of the relevance judgements and the official evaluation metrics of the two \iadh{query sets}, we adopt a different query difficulty classification for TREC 2019 and MSMARCO Dev.
For the TREC 2019 \iadh{query set}, a query is considered \emph{hard}, resp. \emph{easy}, for the BM25 baseline system if the NDCG@10 (the official TREC metric in~\cite{trec19dloverview}) value is \nic{in the first quartile, resp. in the fourth quartile,}
and \emph{medium} otherwise.
%
For the MSMARCO \iadh{query set}, the official metric MRR@10 per query is too sparse to allow percentile computations. Hence we \iadh{consider} a Dev query \iadh{to be} \emph{hard} \iadh{if} its MRR@10 is lesser than or equal to 0.1, and \emph{easy} otherwise.

We partition the queries in each \iadh{query set} according to the corresponding difficulty classification, and compute for how many queries the effectiveness of ANCE and ColBERT is higher (\iadh{denoted} by W(in)) or lower (\iadh{denoted} with L(oss)) than BM25. For each partition, we also compute the average reward and risk associated with \iadh{the} W and L queries, following \cite{10.1145/2348283.2348385}.

\pageenlarge{2}\iadh{Table~\ref{tab:res2} reports the observed results}. For the TREC 2019 queries, both ANCE and ColBERT \craig{exhibit approx.\ the same number of wins/losses for each query difficulty level}.
However, ANCE \iadh{obtains} higher rewards and higher risks on the \iadh{class of easy} queries than ColBERT (+0.1930 vs.\ +0.1827 and -0.1976 vs.\ -0.1380). On the medium difficulty class, the situation \craig{is reversed}, and ColBERT \iadh{obtains} both higher rewards and higher risks than ANCE (+0.3053 vs.\ +02978 and -0.1521 vs.\ 0.1366). On the hard difficulty class, ColBERT is \craig{markedly} superior to ANCE in terms of reward (+0.4114 vs.\ + 0.3750), and risk, even if such risk is computed over a single query.
For \craig{the} MSMARCO Dev queries, ColBERT is able to improve the MRR@10 of both easy and hard queries better than ANCE, and the losses are smaller for ColBERT than for ANCE. 

\looseness -1 \nic{To conclude on RQ2, we have presented experimental evidence \craig{that} both single and multiple representations are approximately as effective on easy queries.} 
\iadh{In contrast, for hard queries,}
the adoption of multiple embeddings helps w.r.t the usage of a single embedding. We explain this by noting that a single \iadh{representation} is learned to compress all semantic information and dependencies of the different tokens composing a query in a single embedding. On the other hand, multiple representations -- using one embedding per query token together with additional masked tokens -- can encode more diverse semantic information in the different embeddings, allowing to retrieve more relevant documents for queries that \iadh{are} hard to answer. 

\begin{table}[tb]
\caption{Comparative \iadh{performances} w.r.t. BM25; queries are classified based on BM25 performance (easy/ medium/hard); Wins and Losses as well as Reward and Risk are calculated w.r.t.\ BM25 performance.}\label{tab:res2}
\begin{adjustbox}{max width=\columnwidth}
\begin{tabular}{lccccc}
\toprule
\multirow{2}{*}{Type} & \multirow{2}{*}{Num} & \multicolumn{2}{c}{ANCE} & \multicolumn{2}{c}{ColBERT} \\
\cmidrule(lr){3-4}\cmidrule(lr){5-6}
        &    &  W/L &  reward/risk & W/L & reward/risk \\
\midrule
\multicolumn{6}{c}{TREC 2019 -- NDCG@10}\\
\midrule
Easy    & 11 &  5/6 & +0.1930/-0.1976 &  5/6 & +0.1827/-0.1380 \\
Medium  & 21 & 17/4 & +0.2978/-0.1366 & 18/3 & +0.3053/-0.1521 \\
Hard    & 11 &  9/1 & +0.3750/-0.1826 & 10/1 & +0.4114/-0.0415 \\
\midrule
\multicolumn{6}{c}{MSMARCO Dev -- MRR@10}\\
\midrule
Easy    & 1954 &  828/712 & +0.4735/-0.4001 & 854/673 & +0.4778/-0.3887 \\
Hard    & 3076 &  1372/24 & +0.4543/-0.1    & 1455/21  & +0.4793/-0.1 \\
\bottomrule
\end{tabular}
\end{adjustbox}
\end{table}

\begin{figure}
\centering
\includegraphics[width=95mm]{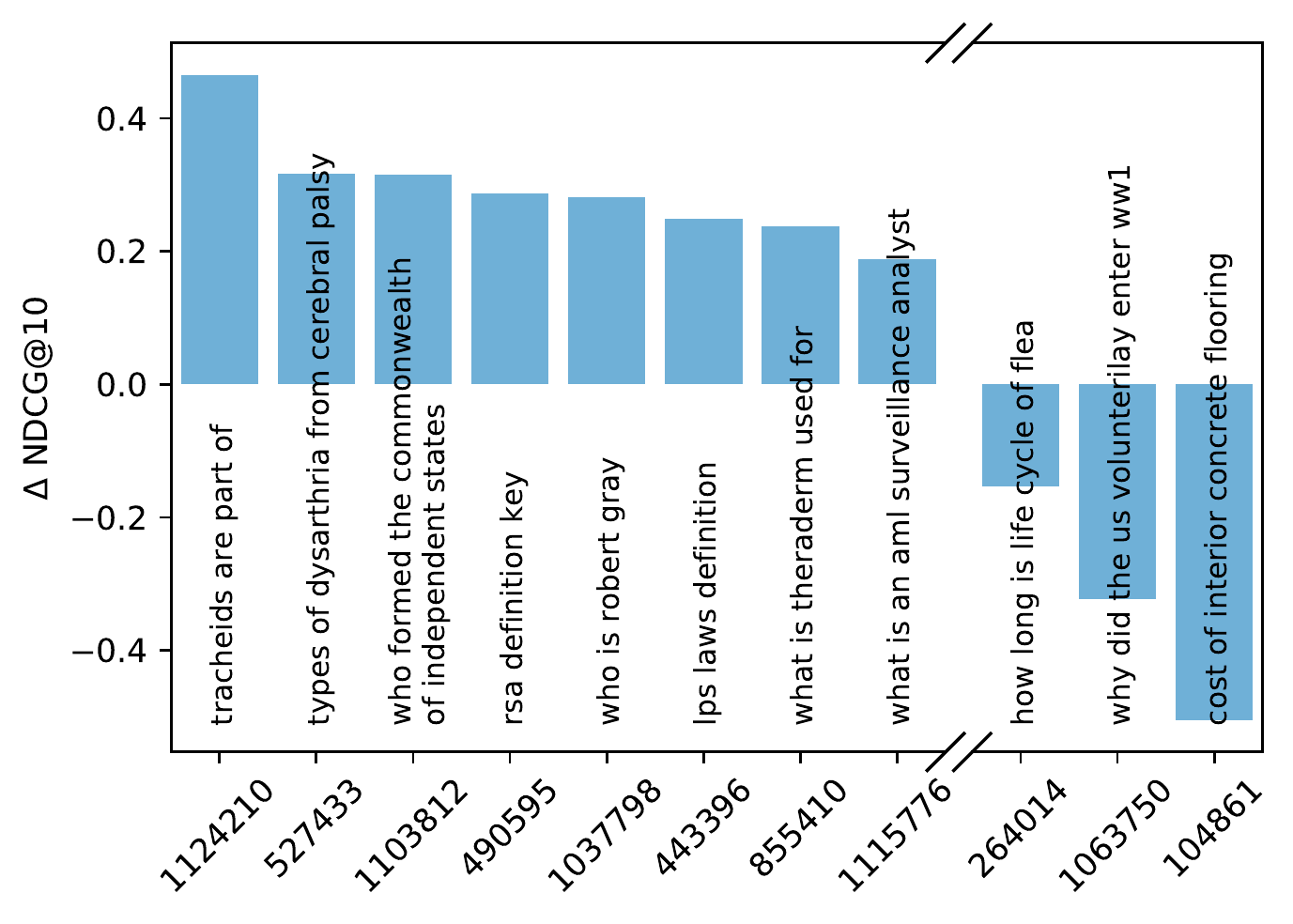}
\caption{Difference in NDCG@10 for queries in the TREC 2019 query set; Differences smaller than 0.15 absolute are omitted. +ve differences are where ColBERT exceeds ANCE. 
}\label{fig:delta}
\end{figure}

\begin{table*}
\caption{Examples of passages retrieved by ANCE and ColBERT at top ranks. The Label column contains the assessment of that document for that query in the qrel file, with -- denoting unjudged.}\label{tab:example}
\scriptsize
\begin{tabular}{|c|cc|p{9cm}|c|}
\toprule
Document & System & Rank & Passage & Label \\
\midrule
\multicolumn{5}{|c|}{ColBERT $>$ ANCE}\\
\midrule
\multicolumn{5}{|c|}{527433: types of dysarthria from cerebral palsy}\\ 
\midrule
8617271 & ColBERT & 1 & There are three major types of dysarthria in cerebral palsy: spastic, dyskinetic (athetosis) and ataxic. Speech impairments in spastic dysarthria involves four major abnormalities of voluntary movement: spasticity, weakness, limited range of motion and slowness of movement. &3\\
8306451 & ANCE & 2 & The types of cerebral palsy are: 1  spastic: the most common type of cerebral palsy; reflexes are exaggerated and muscle movement is stiff. 2  dyskinetic: dyskinetic cerebral palsy is divided into two categories. & 0 \\
\midrule
\multicolumn{5}{|c|}{ANCE $>$ ColBERT}\\
\midrule
\multicolumn{5}{|c|}{1063750: why did the us volunterilay enter ww1}\\ 
\midrule
1300452 & ColBERT & 2 & The main event that led the US to entering ww2 was Japan bombing  Pearl Harbor. The day after the bombing u.s. joined the war   On December 7, 1941, the Japanese Navy lau â.¦ nched a surprise attack  on the naval base at Pearl Harbor, Hawaii.lthough the growing peril of Britain worried many, including Roosevelt, it was not until the US was directly attacked at Pearl Harbor that public and political opinion turned in favor of war with the Axis & -- \\
7952971 & ANCE & 1 & The U.S entered WW1 for several reasons. The U.S entered for two main reasons: one was that the Germans had declared unlimited German submarine warfare and the Zimmermann note.The German had totally disregarded the international laws protecting neutral nation's ships by sinking neutral ships.his note was the last straw, causing Wilson to join the war. The Zimmermann note and unlimited German submarine warfare were two of the biggest cause for the U.S to join the Allies and go to war with Germany. During the war Germany... & 2\\
\bottomrule
\end{tabular}
\end{table*}

\begin{figure*}[tb]
\centering
\includegraphics[width=95mm]{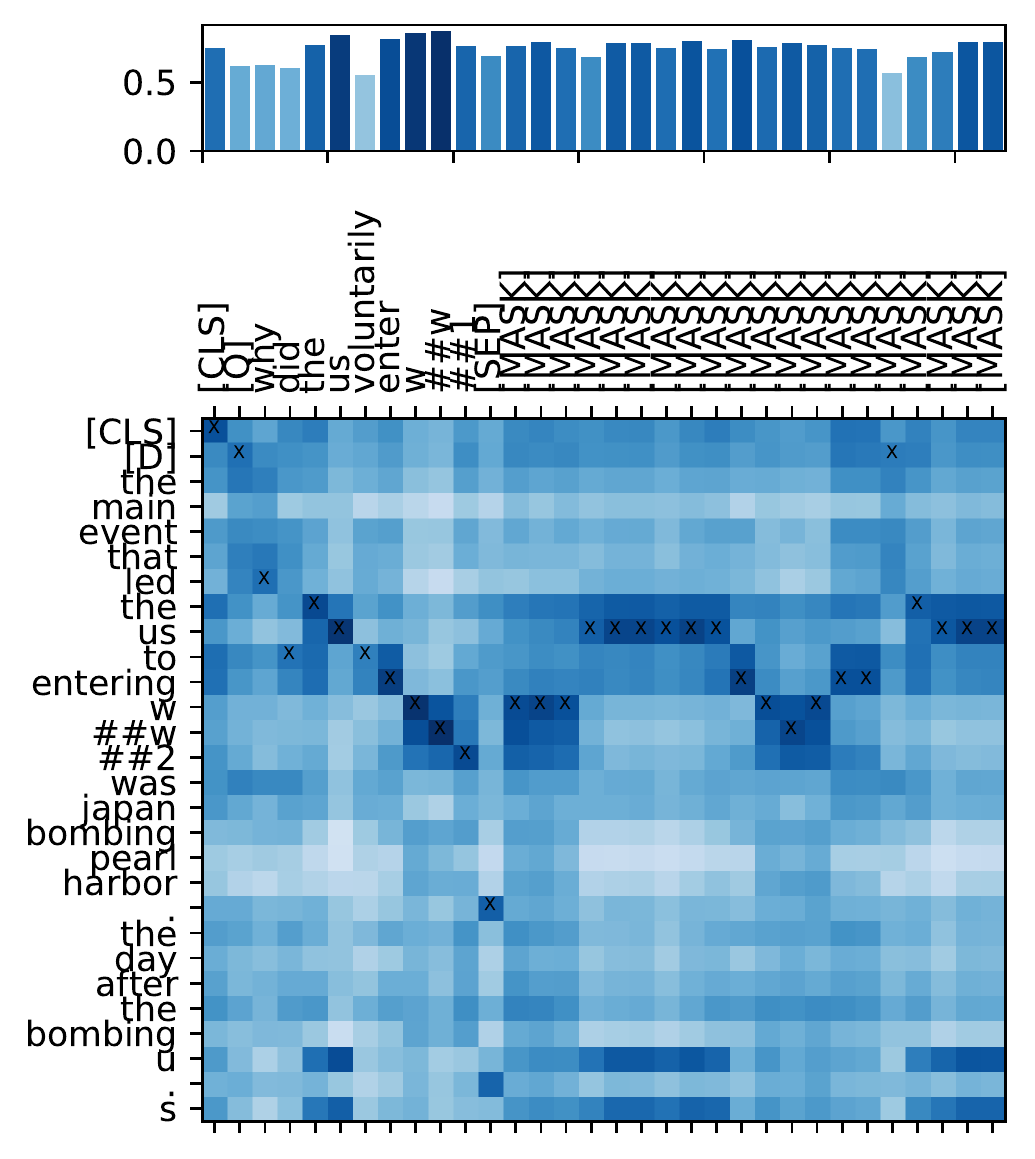}
\caption{\crc{ColBERT interaction between query and document embeddings for query 106375 and passage 1300452 (see Table~\ref{tab:example}). In the interaction matrix, darker shading is indicative of higher similarity; the document embedding (row) with highest similarity for each query embedding (column) is indicated with a $\times$ symbol. The histogram at the top portrays the contribution of each query embedding to the final score of the passage, with shading also indicative of the magnitude of contribution.}}\label{fig:interaction}
\end{figure*}

\subsection{Per-query Comparison}

To address RQ3, in Figure~\ref{fig:delta} we present a per-query histogram comparing the $\Delta$NDCG@10 between ColBERT and ANCE on the TREC 2019 query set; positive deltas indicate a higher NDCG@10 for ColBERT than ANCE. \craig{In total, ColBERT \iadh{outperform} ANCE for 24 queries, while the opposite was true for 17 queries;} $\Delta$s less than 0.15 are omitted for clarity. \craig{On analysing Figure~\ref{fig:delta}, it appears that many queries requesting a \textit{definition} appear to perform well for ColBERT (e.g. 1124210, 490595). Indeed, on closer inspection of the TREC 2019 query set, out of 43 queries, we found 19 such definitional queries -- of which 16 were more effective for ColBERT.}

To illustrate \craig{other} differences between the approaches, in Table~\ref{tab:example} we select \craig{two non-definitional} queries where one approach markedly outperformed the other (but not the queries with the most extreme deltas, which may be outliers). Firstly, for query 527433 (`types of dysarthria from cerebral palsy'), ColBERT identifies a passage that clearly answers the query; in contrast, the non-relevant passage identified at rank 2 by ANCE appears to have focused \craig{solely} on the `cerebral palsy' aspect, omitting the dysarthria aspect \craig{of the query}. Indeed, the Precision @10 of ANCE for this query was $\frac{3}{10}$, \craig{compared to  $\frac{6}{10}$ for ColBERT}. This suggests that ANCE's compression of a complex information need into one embedding has caused an information loss, with the model focusing on only a single aspect of the query, resulting in low effectiveness.

\pageenlarge{1} On the other hand, for query 1063750 (`why did the us volunterilay enter ww1'), ANCE identified a relevant passage, but ColBERT identified a passage (1300452) focusing entirely on the wrong World War (`ww2' rather than `ww1'). %
%
At least some of the reason for the conflation of meanings is that \craig{neither} `ww1' nor `ww2' do not appear in BERT's fixed vocabulary, e.g., the latter is tokenised into word pieces as `w', `\#\#w', `\#\#2'. Hence distinguishing between `ww1' and `ww2' information needs require context to be \craig{distributed} across the three embeddings. \crc{To analyse this passage further, Figure~\ref{fig:interaction} shows the ColBERT interaction between the query and document embeddings for this passage and query\footnote{\crc{This figure can be reproduced using the {\tt explain\_text()} function within our PyTerrier\_ColBERT library.}}. In the figure, the darker shading in the matrix is indicative of higher similarity; the highest similarity that is selected for a given query embedding by the max-sim operator is indicated by a $\times$ symbol; the histogram at the top of the figure indicates the contribution of each query embedding to the final passage score. Indeed, on inspection of the max similarities for this passage shows that} the highest contributions to the passage's score comes from the `\#\#w' token, with `\#\#1' query embedding \craig{being highly similar to the} `\#\#2' document embedding. This suggests that the embeddings for `\#\#1' and `\#\#2' are not sufficiently contextualised when following `\#\#w', or that ColBERT's max similarity computation could be adapted to better address proximity. In contrast, ANCE retrieved passage 1300452 at rank 155, \craig{showing that the single representations for the passages sufficiently distinguish between World War 1 vs.\ World War 2.}


In summary, in addressing RQ3, we observed that there \iadh{exists} some \iadh{large differences} between ANCE and ColBERT for some queries. Our analysis found that ColBERT perfoms better than ANCE for definitional type queries. Moreover, our analysis suggests that in ANCE, the use of a single embedding representation risks misinterpreting complex queries with multiple aspects \nic{as shown by results in the previous subsection}; For ColBERT, the max similarity operator can overly focus on highly similar embeddings \iadh{at the} risk \iadh{of} mis-interpreting a query.
\vspace{-\baselineskip}
\pageenlarge{1}\section{Conclusions}
\looseness -1 Despite their recency, dense passage retrieval approaches have the effectiveness potential to supplant the traditional inverted index data structure. Yet, different families of dense retrieval are emerging, for which the comparative advantages and disadvantages are not yet clear. In this work, we made a systematic study of single vs.\ multiple representation dense retrieval approaches, namely ANCE and ColBERT.  We found that while both significantly outperformed BM25 baselines across various metrics, ColBERT significantly outperformed ANCE for MAP on TREC2019 and MRR@10 on the MSMARCO Dev query set, was more effective for queries that BM25 found hard, and was better at definitional queries as well as queries that had complex information needs. On the other hand, ANCE has desirable qualities in terms of mean response time and memory occupancy \craigi{(see Table~\ref{tab:summary})}. We postulate that research should be directed toward hybrid solutions, either reducing the size of the ColBERT embedding index, e.g., through adaptations to static pruning, or through using multiple embeddings within ANCE for complex queries/passages.



 
\section*{Acknowledgements}

Nicola Tonellotto was partially supported by the Italian Ministry of Education and Research (MIUR) in the framework of the CrossLab project (Departments of Excellence). Craig Macdonald and Iadh Ounis acknowledge EPSRC grant EP/ R018634/1: Closed-Loop Data Science for Complex, Computationally- \& Data-Intensive Analytics. 

\bibliography{bib}\pageenlarge{1}
\end{document}